\documentclass[final]{elsarticle}

\usepackage{graphicx}
\usepackage{graphicx,psfrag,epsf,epsfig}
\usepackage{bm,bbm}
\usepackage{pstricks,pst-plot}     
\usepackage{latexsym}
\usepackage{mathrsfs,amsmath,amssymb,textcomp}
\usepackage[figuresright]{rotating}

\journal{Solar Energy Materials \& Solar Cells}

\begin{document}

\begin{frontmatter}

\title{Spectral properties of photogenerated carriers in quantum well
solar cells}

\author{U. Aeberhard}

\address{IEF-5 Photovoltaik, Forschungszentrum
J\"ulich, D-52425 J\"ulich, Germany}

\begin{abstract}
The use of low-dimensional structures such as quantum wells, wires or dots in
the absorbing regions of solar cells strongly affects the spectral response of
the latter, the spectral properties being drastically modified by quantum
confinement effects. Due to the microscopic nature of these effects, a
microscopic theory of absorption and transport is required for their
quantification. Such a theory can be developed in the framework of the
non-equilibrium Green's function approach to semiconductor quantum transport and
quantum optics. In this paper, the theory is used to numerically investigate
the density of states, non-equilibrium occupation and corresponding excess
concentration of both electrons and holes in single quantum well structures embedded 
in the intrinsic region of a $p$-$i$-$n$
semiconductor diode, under illumination with monochromatic light of different energies. Escape of
carriers from the quantum well is considered via the inspection of the
spectral photocurrent at a given excitation energy. The investigation shows
that escape from deep levels may be inefficient even at room temperature. 
\end{abstract}


\end{frontmatter}

\section{Introduction}

Efficient carrier escape from the quantum wells is one of the main premises for
high performance of quantum well solar cells (QWSC), since for a contribution to
photocurrent, the carrier sweep-out rate must exceed the recombination rate
in the well. An advanced understanding of the escape mechanisms is thus
indispensible for further optimization of such devices. To study carrier escape
processes in QWSC, combined temperature and bias dependent photoluminescence and 
photocurrent measurements on $p$-$i$-$n$ diodes including single quantum wells (SQW)
in the intrinsic region were performed
\cite{Nels93,Thuc96,Barn97,Zach98,Mcfa99}. 
From the experimental results, two principle escape mechanisms were
identified: thermal emission over \cite{Schn88} and tunneling through the confining barrier, where the
latter can be assisted thermally or by scattering with phonons \cite{Lars88} or ionized
impurities \cite{Fox91}. Both mechanisms depend on material properties and
design parameters like width and height of the barriers as well as
on external conditions like terminal voltage and temperature. In the case of
thermionic emission, the escape rate increases with lower barrier and higher temperature, 
where the field dependence enters via the barrier height. 
Tunneling escape is determined via the quantum-mechanical transmission through
the finite height barrier and increases with decreasing barrier thickness, as a 
consequence of weaker confinement that leads to
broader levels corresponding to a shorter lifetime, and growing 
fields enabling Fowler-Nordheim tunneling by lowering the
effective barrier height.
At very high fields corresponding to reverse bias condition, carrier escape is 
found to be maximum, which is explained by the effective barrier being both
narrow and low. At low fields corresponding to forward bias, i.e. for the situation that is
relevant for photovoltaic operation,
three temperature regimes with specific prevailing carrier escape channels
have been established. At very low temperature, tunneling dominates, since on
the one hand, there is no significant thermal population of higher levels near the top of the well
that would allow thermal emission,
and on the other hand the coherence length is increased. At intermediate
temperature (100-200 K), tunneling becomes thermally assisted, via phonons or
ionized impurity scattering. Finally, at the high (room) temperatures, escape was found to be dominated by thermionic emission,
leading in many cases to an internal quantum efficiency close to unity \cite{Nels93,Yaza94}.

The experimental investigation of carrier escape from SQW was supplemented by a
number of modelling approaches with increasing complexity, based on semiclassical pictures of
thermionic emission and quantum mechanical descriptions of tunneling escape in
terms of the barrier transmission function, which together allowed the
determination of a carrier escape rate \cite{Nels93,Lars88,Barn94,Moss94}.
However, these models share a set of shortcomings that have not been adressed properly since
then. For instance, approximate models for the density of states were used, with
abrupt dimensionality transition at the top of the well or even neglecting the lower dimensionality
of the quantum well states. The confinement levels were determined without proper 
consideration of the quantum well in-plane, or transverse, dispersion. For the
occupation of the confinement levels, the quasi-Fermi level was chosen to lie on 
the lowest level in the well, which is only true at large enough forward bias,
and was not determined in function of the photoexcitation characteristics. Moreover, 
even though the effects of electron-phonon
scattering on tunneling and occupation of higher-lying states in the well were
considered in a few cases on a microscopic basis, they were not incorporated
in a transport model beyond the rate equation level.

In this paper, we adress the issue of photocurrent generation in SQW
$p$-$i$-$n$ diodes on the basis of the Non-equilibrium Green's function (NEGF)
formalism, a microscopic quantum-kinetic theory dating back to Keldysh
\cite{Keld65}, Kadanoff and Baym \cite{Kada62}. Within the framework of
this theory, many of the limiting assumptions inherent to the quasiclassical
Boltzmann picture can be relaxed, and energy resolved information is obtained
for most of the relevant quantities. It allows for a combination of quantum
mechanical descriptions of the electronic structure and the interactions of the
many-body system composed of electrons, holes, photons and phonons with a
theory of transport in open quantum systems, including both coherent
and incoherent transport processes, and is therefore well suited for the
investigation of the problem at hand where electron-hole pair photogeneration
and carrier-phonon scattering have to be treated on equal footing with tunneling
transport.

The paper is organized as follows. After a brief review of the
theoretical model in the following section, we present and discuss the
numerical results for the excess carrier density, the corresponding occupation
function and the local photocurrent spectrum for different excitation energies in 
the range of the quantum well absorption.

\section{Microscopic model of QWSC} 

The analysis in the present work relies on the microscopic theory of quantum
well solar cells as discussed in detail in Ref. \cite{Aebe08}, which is based
on the steady state non-equilibrium Green's function formalism. Since within
this framework, the determination of the optical and electronic properties of a
device differ considerably from the semiclassical macroscopic continuum
approach conventionally used in photovoltaics, a brief sketch of the method
shall be presented here.

\subsection{The NEGF formalism}

In the approach under discussion, physical quantities of interest, such as
spectral density or current, are determined via the calculation of the
steady-state non-equilibrium Green's functions for the charge carriers. These
quantities are defined as non-equilibrium quantum statistical ensemble averages
of pairs of single particle field operators. The retarded Green's function
$G^{R}(\mathbf{r},t;\mathbf{r'},t')$, which provides e.g.
the local density of states, and the correlation functions $G^{\lessgtr}(\mathbf{r},t;\mathbf{r'},t')$ 
providing density and current for electrons and holes, respectively, are obtained via the solution of 
the corresponding equations of motion, which for steady state ($t-t'\rightarrow E)$ read\footnote{We neglect the 
spin degrees of freedoms in the present discussion up to consideration of a
factor of 2 corresponding to the sum over degenerate spin components.}
\begin{align}
 G^{R(A)}({\mathbf r_{1}},{\mathbf
 r}_{1'};E)=&G_{0}^{R(A)}({\mathbf r_{1}},{\mathbf
 r}_{1'};E)+\int d^{3}r_{2}\int
 d^{3}r_{3}G_{0}^{R(A)}({\mathbf r}_{1},
 {\mathbf r}_{2};E)\nonumber\\&\times\Sigma^{R(A)}({\mathbf r}_{2},{\mathbf
 r}_{3};E) G^{R(A)}({\mathbf r}_{3},{\mathbf
 r}_{1'};E),\label{eq:dyson1}\\
 G^{\lessgtr}({\mathbf r}_{1},{\mathbf
 r}_{1'};E)=&\int d^{3}r_{2}\int d^{3}r_{3} 
 G^{R}({\mathbf
 r}_{1}, {\mathbf r}_{2};E)\Sigma^{\lessgtr}({\mathbf
 r}_{2},{\mathbf r}_{3};E) G^{A}({\mathbf
 r}_{3},{\mathbf r}_{1'};E),\label{eq:dyson2}
 \end{align}
with the non-interacting retarded (advanced) Green's function given by
\begin{equation}
\left\{G_{0}^{R(A)}\right\}^{-1}(\mathbf{r},\mathbf{r}',E)=\left[E+(-)i\eta-H_{0}(\mathbf{r})-U(\mathbf{r})\right]
\delta(\mathbf{r}-\mathbf{r'}),\quad\eta\rightarrow 0^{+},
\end{equation}
where $H_{0}$ is the Hamiltonian of the non-interacting system and $\eta$
provides the correct analytical properties. $\Sigma^{R,\lessgtr}$ denote the
retarded self-energy and the scattering functions, respectively, and are
composed of two contributions, due to the contacts and due to interactions
within the active device. The effects of open boundaries and carrier injection
from contact reservoirs are included via the boundary self-energies
$\Sigma^{R,\lessgtr}_{B}$, which are determined by the bulk Bloch states of the
electrodes that are occupied according to the chemical potential of the contact. 
The interaction of electrons and holes with phonons and photons is considered perturbatively 
via corresponding self-energy terms $\Sigma^{R,\lessgtr}_{int}$. Concerning the matter-light coupling, the
dipole approximation for a monochromatic single-mode photon field is used for
the optical excitation, and the coupling to a continuum for isotropic radiative
emission. Elastic scattering by acoustic phonons is included in the
deformation potential formulation, and the inelastic coupling to polar
optical phonons is described via the Fr\"ohlich Hamiltonian of the harmonic
approximation. 

Together with the expressions for the self energies from boundaries and
interactions, which are calculated exactly (boundaries)
or perturbatively within self-consistent first Born approximation
(interactions) and depend linearly on the carrier Green's functions, and the
macroscopic Poisson equation
\begin{equation}
\epsilon_{0}\nabla_{\mathbf{r}}\left[\epsilon(\mathbf{r})\nabla_{\mathbf{r}}U(\mathbf{r})\right]=
n(\mathbf{r})-p(\mathbf{r})-N_{dop}(\mathbf{r}),
\label{eq:poisseq}
\end{equation}
relating the Hartree potential $U$ to doping density $N_{dop}$ and
the carrier densities derived from the Green's functions, relations
\eqref{eq:dyson1} and \eqref{eq:dyson2} form a closed set of equations for the
latter that have to be solved self-consistently.

\subsection{Determination of physical quantities form NEGF}

For the numerical investigation of quasi one-dimensional multilayer-systems such
as multi quantum wells and superlattices, the Hamiltonian is represented in a
discrete basis, e.g. consisting of a suitable combination of localized atomic
orbitals $|\alpha,L,\mathbf{k}\rangle$, where $\alpha$ denotes the
atomic orbitals, $L$ the model layer and $\mathbf{k}$ the in-plane, or transverse,
momentum. In this basis, the local density of states (LDOS) at layer $L$ is
given by
\begin{align}
\mathscr{D}_{L}(E)&=\sum_{{\mathbf k}}\mathrm{tr}\{A_{L;L}({\mathbf
k};E)\},\qquad A_{L;L}({\mathbf
k};E)=i\left[G^{R}_{L;L}({\mathbf
k};E)-G^{A}_{L;L}({\mathbf
k};E)\right],
\end{align}
where $A$ is the spectral function and the trace is over orbital indices. The averaged electron (hole) density 
at layer $L$ is
\begin{equation}
  n_{L}=-\frac{2i}{\mathcal{A}\Delta}\sum_{{\mathbf
 k}}\int\frac{dE}{2\pi}tr\{G^{<}_{L;L}({\mathbf k};E)\},\qquad
 p_{L}=\frac{2i}{\mathcal{A}\Delta}\sum_{{\mathbf
 k}}\int\frac{dE}{2\pi}tr\{G^{>}_{L;L}({\mathbf k};E)\},
\end{equation}
where $\mathcal{A}$ denotes the cross section area and $\Delta$ the layer thickness.
The current density passing from layers
$L$ to $L+1$ is 
\begin{align}
 J_{L,L+1}^{n(p)}&=\frac{2e}{\hbar \mathcal{A}}\sum_{{\mathbf
 k}}\int\frac{dE}{2\pi}tr\{t_{L;L+1}G^{<(>)}_{L+1;L} ({\mathbf k};E)-t_{L+1;L}G^{<(>)}_{L;L+1}({\mathbf k};E)\}\label{eq:current}.
\end{align}
It is possible to define an effective local carrier
distribution function $\tilde{f}_{L}(E)$ via
\begin{equation}
\tilde{f}_{L}(E)\equiv\frac{\rho_{L}(E)}{\mathscr{D}_{L}(E)}=\frac{\sum_{\mathbf{k}}
\mathrm{tr}\{-iG^{<}_{L;L}(\mathbf{k};E)\}}{\sum_{\mathbf{k}}\mathrm{tr}
\{A_{L;L}(\mathbf{k};E)\}},\label{eq:occ}
\end{equation}
where $\rho_{L}(E)$ is the spectral density at
layer $L$.

\section{Numerical results}

The following results for bulk and SQW $p$-$i$-$n$-diodes were all obtained
using the two-band tight-binding $sp_{z}$-Hamiltonian discussed in Ref.
\cite{Aebe08}, with the same set of parameters corresponding to a
GaAs-Al$_{x}$Ga$_{1-x}$As heterostructure, but with energy gaps of 0.5 eV
(well) and 0.9 eV (barrier), and band offset (=barrier
height) of 0.25 eV in the conduction band and 0.25 eV in the valence band. The
active device region, where interactions are considered, includes a SQW of 25 monolayers (ML) 
width and 5 ML adjacent barrier material.
The contacts are formed by 50 ML high bandgap material with strong
doping ($N_{d,a}=10^{18}$ cm$^{-3}$). Between contact and active device, intrinsic buffer 
regions of 60 ML are inserted. The
calculations are performed at 300 K and an illumination intensity of 1 kW/m$^{2}$.

\subsection{Photogenerated excess carrier density}

\begin{figure}[b!]
\begin{center}
\includegraphics[width=12cm]{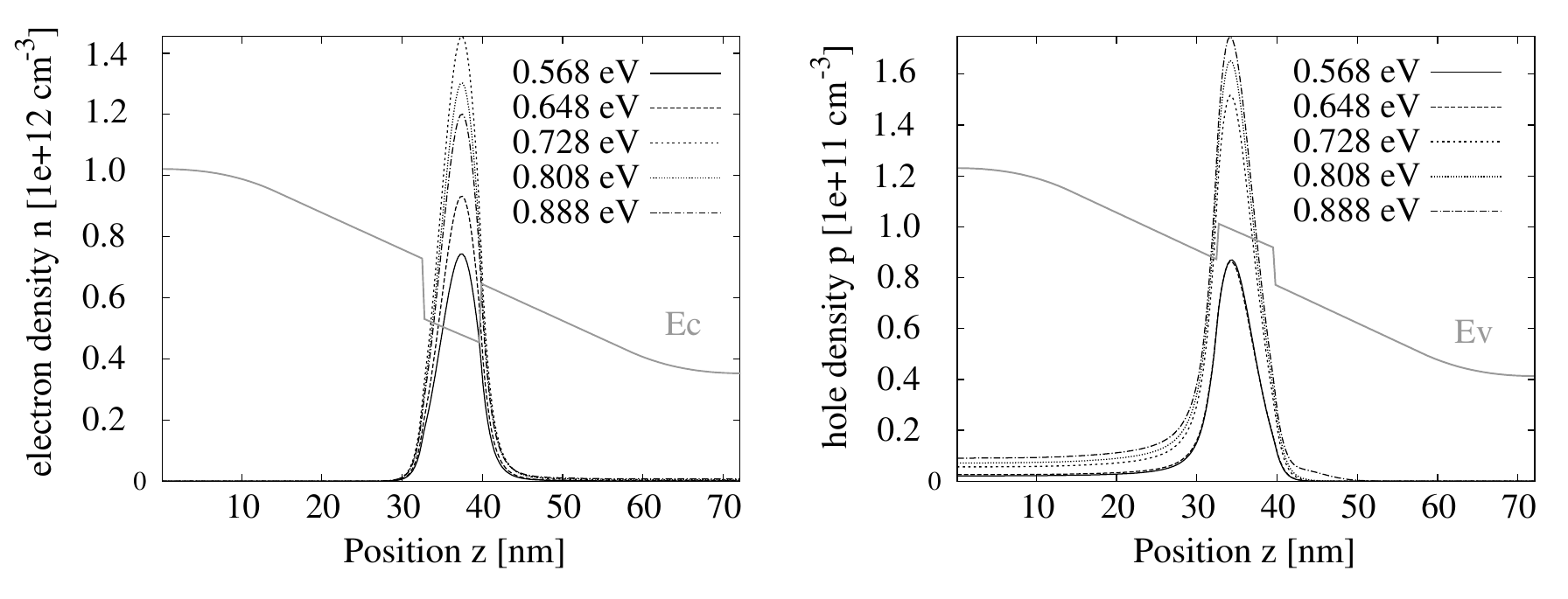}
\caption{Photogenerated excess carrier density
for different photon energies at $V_{bias}=-0.01V$. The quantum well location
is indicated by the band edges $E_{C}$ and
$E_{V}$, respectively. The electronic density exceeds the hole concentration
due to stronger confinement and resulting slower carrier escape. While the electron density is
 largest at intermediate excitation energies due to the absence of efficient
 escape channels, the increase of the hole density closely follows the absorption, which at low energies
grows step-like.
\label{fig:dens_light_sqw}}
\end{center} 
\end{figure} 
Fig. \ref{fig:dens_light_sqw} shows the photogenerated excess carrier density
$\delta n=n_{light}-n_{dark}$ for the SQW diode and several photon energies
lying in the range of the confinement level separation between the two band gap
values, such that the contact and lead regions are non-absorbing. 

Owing to the localized nature of the states which are occupied, the
fraction of the carriers that leaks out of the well is small,
especially in the case of the electrons, which are more confined than the
holes. The higher electron concentration is a consequence of the longer escape 
times of the electrons resulting from this stronger confinement. It is further 
interesting to notice that while
the hole concentration follows the increase in photocurrent with larger photon energy,
this is not the case for the electron concentration. This observation is
explained by the stronger increase in escape channels for electrons at higher
excitation energies as compared to the situation of the holes, where the
enhancement sets in at lower energies, i.e. the slow escape of electrons at
intermediate excitation energies leads to charge build-up in the well.

\begin{figure}[t!]
\begin{center}
\begin{minipage}{5.5cm} 
\includegraphics[width=5.5cm]{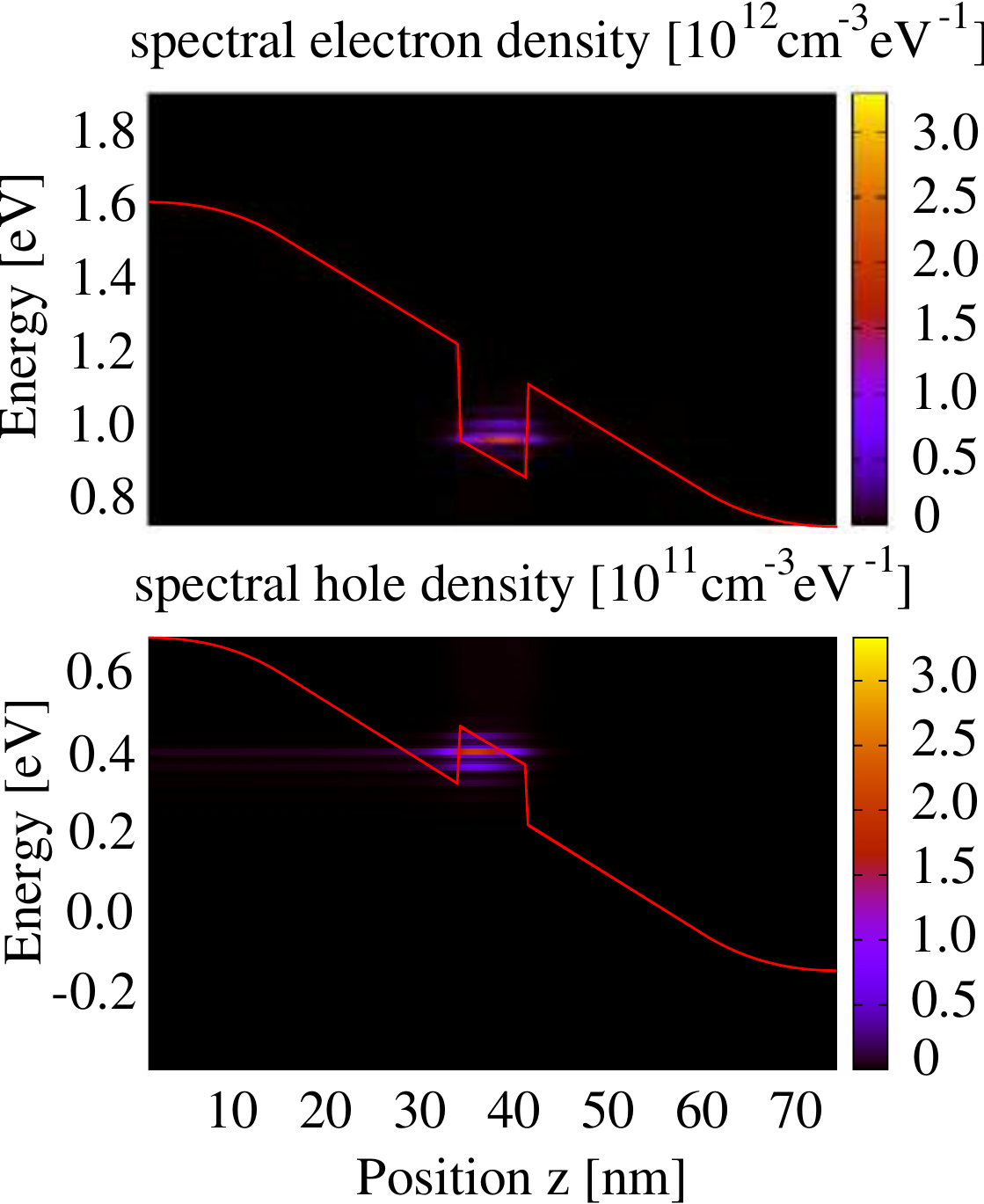}
\end{minipage}\qquad
\begin{minipage}{5.5cm}
\includegraphics[width=5.5cm]{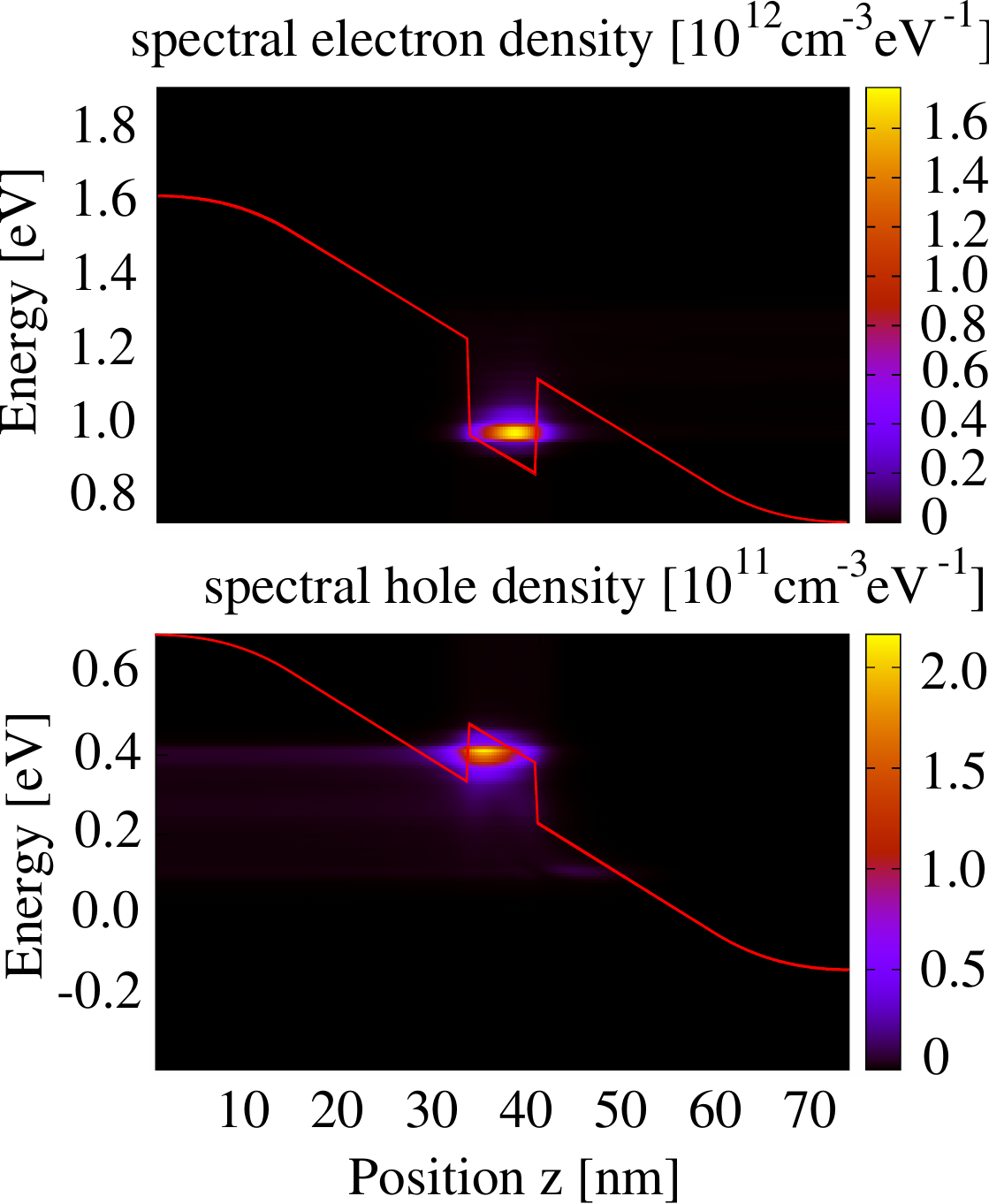}
\end{minipage} 
\begin{minipage}{5.5cm}
\caption{(Color online) Photogenerated excess carrier spectral density
for illumination with $E_{phot}=0.568~eV$ (bound-state transition), at
$V_{bias}=-0.01$ V. Interaction with phonons leads to the formation of
distinct satellites peaking at the energy distance of an integer multiple of the
phonon energy $\hbar\omega_{LO}=0.036$ eV. \label{fig:diffoc_sqw_568}}
\end{minipage}\qquad
\begin{minipage}{5.5cm} 
\caption{(Color online) Photogenerated excess carrier spectral density
for illumination with $E_{phot}=0.888~eV$ (quasi-continuum
transition), at $V_{bias}=-0.01$ V. The phonon peaks are no longer
distinguishable due to larger broadening of the higher, less confined
states which are occupied at this excitation energy.
 \label{fig:diffoc_sqw_888}}
 \end{minipage}
\end{center}
\end{figure}

\subsection{Non-equilibrium occupation of confinement levels}
To investigate the occupation of available states by the photogenerated excess
carriers, the spectral density $\delta\rho_{L}(E)$ of the latter is
computed for different excitation energies. In Fig. \ref{fig:diffoc_sqw_568}, the photogenerated excess carrier
spectral density for illumination with $E_{phot}=0.568~eV$ (bound state transition) is
displayed. At this low energy, the photogenerated carriers occupy only the
lowest subband. This means that not only the dark
carrier concentration, but also the photogenerated carrier
density is strongly localized in energy around the lowest
level, and the occupation of higher and lower energy states 
via phonons produces pronounced satellite peaks, both in
the conduction band well and the valence band well.

\begin{figure}[t!] 
\begin{center}
\includegraphics[width=12cm]{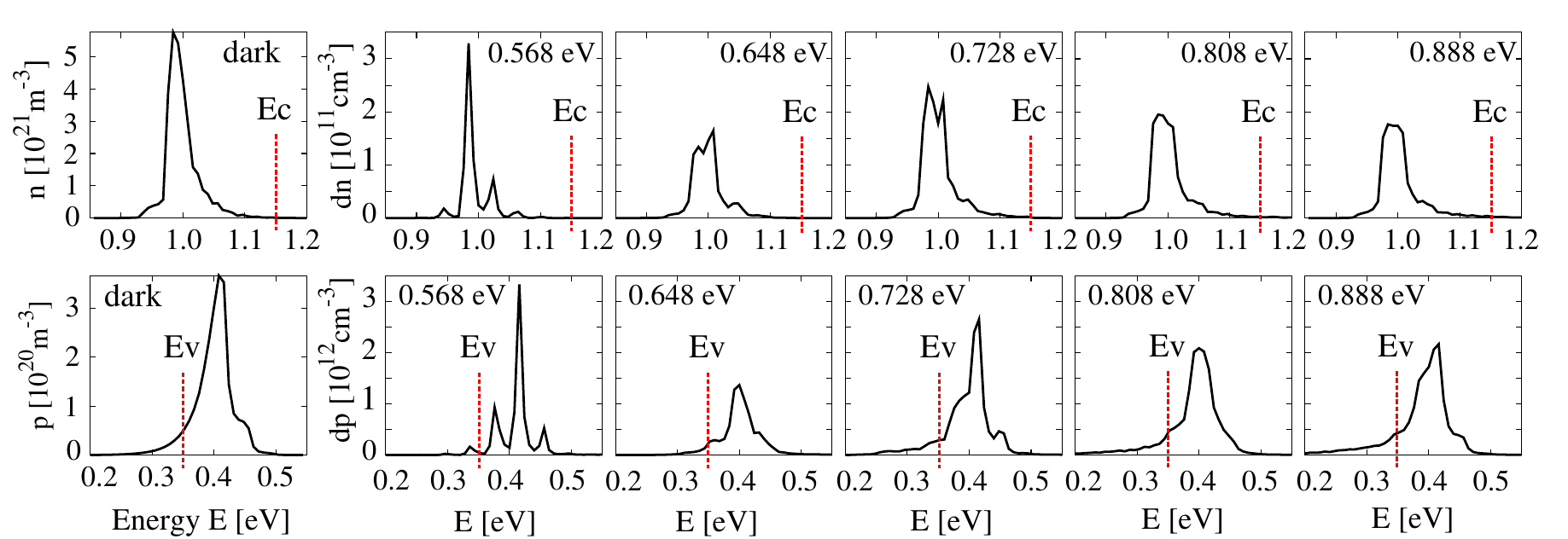}
\caption{Spectral resolution of the excess carrier density
for different photon energies (cut at the position of the
maximum in the quantum well). For comparison, the
carrier concentration in the dark is given for the
same position in the quantum well. $E_{C}$ and $E_{V}$ indicate the effective
well edges for electrons and holes, respectively. While the phonon satellite
peaks get smoothened at higher excitation energies due to occupation of broadened states, 
the highest weight remains always on the lowest confinement level. Only the
hole density shows a significant fraction at energies above the effective
well edge $E_{V}$, which is required for efficient thermionic emission.
\label{fig:spectdens_light_sqw}}
\end{center} 
\end{figure}  
\begin{figure}[t!]   
  \begin{center} 
   \includegraphics[width=10cm]{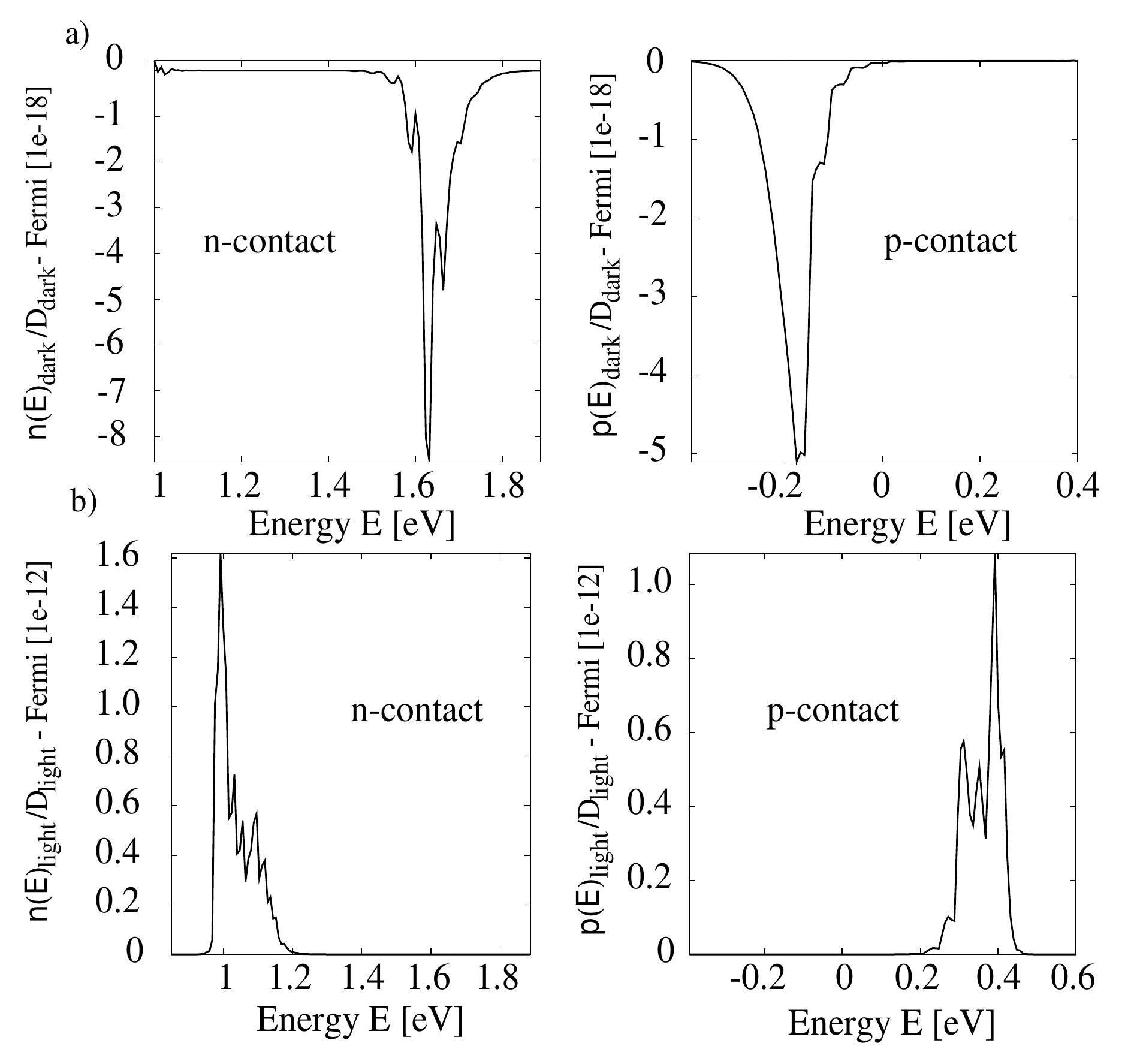}
    \caption{Distribution function
    $\tilde{f}_{L}(E)\equiv\rho_{L}(E)/\mathscr{D}_{L}(E)$ at the majority
    carrier contact ($L=N_{z}$ for electrons, $L=1$ for holes) under small
    forward bias ($V=-0.01~V$), in the dark and under illumination, with
    comparison to the equilibrium Fermi distribution $f_{\mu}(E)$ of the
    contacts with chemical potential $\mu$:  (a) deviation $\delta
    f=\tilde{f}-f_{\mu}$ from the equilibrium distribution in the dark, (b)
    deviation from equilibrium under illumination.} 
    \label{fig:occupation}  
  \end{center}  
\end{figure} 
 
The photogenerated excess carrier spectral density
for illumination with $E_{phot}=0.888~eV$ (quasi-bound state transition) is
shown in Fig. \ref{fig:diffoc_sqw_888}. At this photon energy, states high in
the quantum well that are broad due to fast escape are occupied by the
photoexcitation process. Subsequent carrier relaxation via emission of phonons leads to the occupation
of the lower confinement levels, which then provide the main contribution to the spectral density. 
Due to the width of the states of initial occupation, the final excess carrier
spectrum exhibits only smooth features. 
This smoothing of the phonon satellite peaks with increasing excitation energy
and the accumulation of density on the lowest confinement levels can also be seen in Fig.
\ref{fig:spectdens_light_sqw}, which shows a cut of the spectral excess carrier
concentration at the position of its maximum in the well. What is interesting
to notice in this figure is the large difference in the degree of thermionic
emission between electrons and holes: while the excess carrier density
in the valence band has a tail that extends far above the effective barrier edge
at $E_{V}\sim 0.35 ~eV$, the corresponding density is very low in conduction
band well at the effective barrier edge of $E_{C}\sim 1.15 ~eV$, even at the
highest excitation energies. The phonon satellite peaks get smoothed at higher
excitation energies due to occupation of broadened states, but the highest
weight remains always on the lowest confinement level, independent of the energy of the
photon.

The occupation function corresponding to the excess carrier density is
evaluated using relation \eqref{eq:occ}. Fig. \ref{fig:occupation} shows this
distribution at the majority carrier contact ($L=N_{z}$ for electrons, $L=1$ for
holes) under small forward bias ($V=-0.01~V$), in the dark and under illumination, 
with comparison to the equilibrium Fermi distribution $f_{\mu}(E)$ of the
contacts with chemical potential $\mu$. In the dark, the only
evidence of non-equilibrium giving rise to a deviation $\delta
f=\tilde{f}-f_{\mu}$ from the equilibrium distribution is the signature of the
leakage current from the opposite contact, which is characterized by a chemical
potential $\mu'$ that differs from $\mu$ by the applied bias voltage,
$\mu'=\mu+V_{bias}$. Under illumination, there is an additional deviation from equilibrium which 
reflects the distribution of the photogenerated excess carriers.

\subsection{Photocurrent spectrum and carrier escape from quantum well states}

\begin{figure}[h!]
\begin{center} 
\includegraphics[width=12cm]{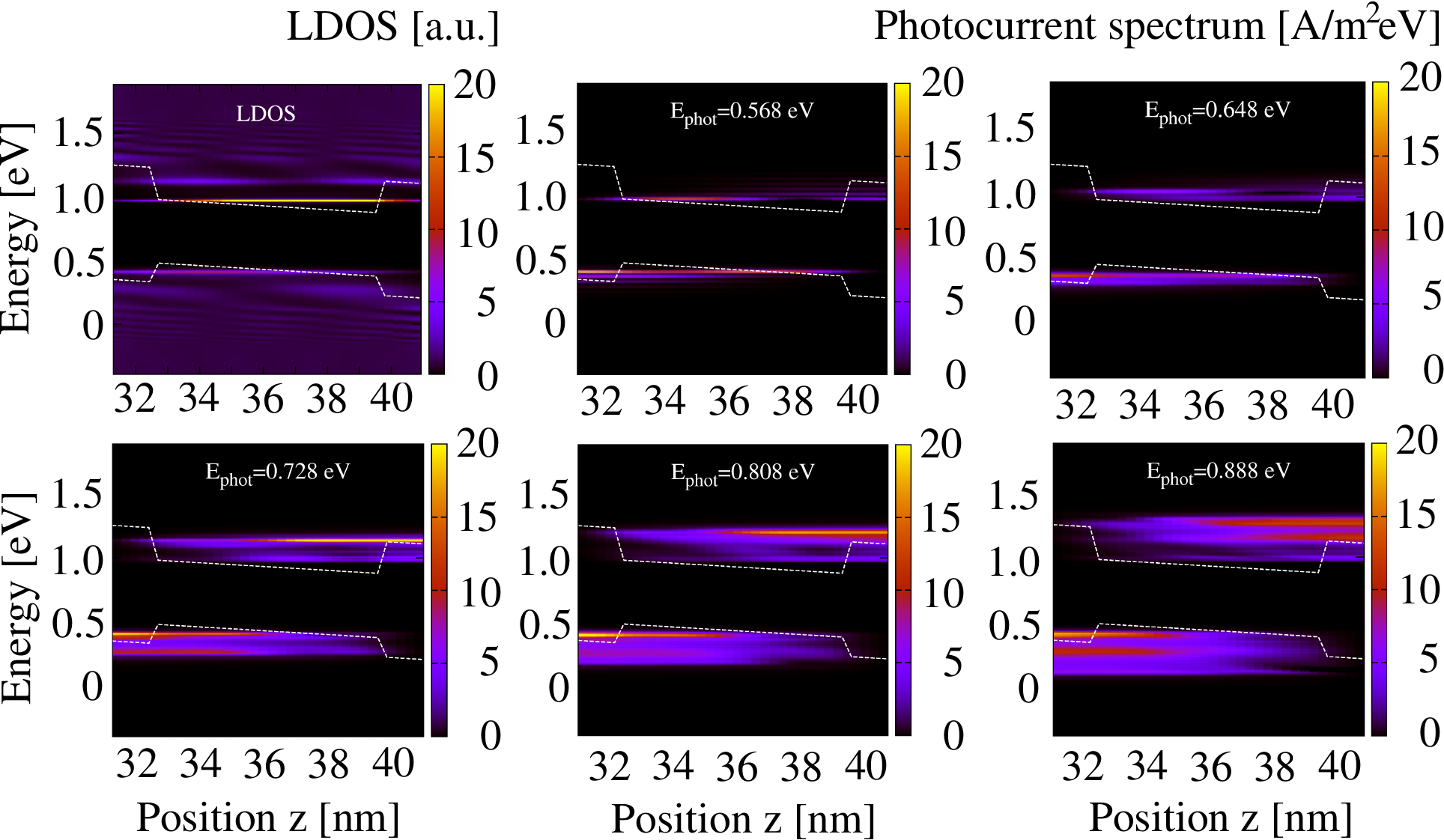}
\caption{(Color online) Local density of states and energy resolved local
photocurrent at $V_{bias}=-0.01~V$, illustrating the carrier escape channels at different
photon energies. Due to the presence of a high field, tunneling
escape is possible even for the lower levels. At low excitation energy,
transport is restricted to tunneling, whereas thermionic and phonon-assisted
emission set in with the occupation of higher subbands.
\label{fig:ixe} }
\end{center}
\end{figure}
Fig. \ref{fig:ixe} shows the local density of states and energy resolved local
photocurrent in the active device region at $V_{bias}=-0.01~V$, illustrating the
carrier escape channels at different photon energies. At $E_{phot}=0.568~eV$,
only the lowest subbands are occupied (see also Fig. \ref{fig:diffoc_sqw_568}).
In the case of such a short structure and at low forward bias, escape from this level is
possible via field enhanced tunneling. At $E_{phot}=0.648~eV$, the occupation is
increased, but transport is still restricted to the lowest levels. This means
that thermionic emission is not an efficient escape channel for deep
levels, and the emission via phonon absorption is limited by the large
separation of the subbands. At $E_{phot}=0.728~eV$, the occupation of the higher
subbands has set in. From there, escape is efficient since the states are no longer 
strictly confined to the well, and thermionic emission is possible. The high level current increases
further at $E_{phot}=0.808~eV$, and at $E_{phot}=0.888~eV$, additional
quasi-bound states have started to contribute. The lower levels still contribute, but only a part of the subband
carries current. An important result of this investigation is the observation
that in the present case of very high fields, where tunneling from low levels is
possible, this channel completely dominates the carrier escape, i.e. the contribution of thermionic
emission is negligible in comparison. 

\section{Summary and conclusions}

We have numerically investigated photocurrent generation in SQW $p$-$i$-$n$
diodes using the NEGF-formalism, taking into account the correct density of
states in the well for a two band model with parabolic transverse dispersion,
the tunneling through the confining barrier and the scattering with acoustic
and polar optical phonons. The occupation of the confined states is determined
in function of external conditions such as energy and intensity of the
photoexcitation and the separation of the contact Fermi-levels corresponding to
a bias voltage. The computed excess carrier density strongly depends on the
the excitation energy and thus on the level structure of the well. At the large
fields that were investigated, tunneling is very efficient and represents the
dominant escape channel. At low photon energies, where only the lowest subbands
are partially populated, thermal occupation of higher states does not
contribute significantly to the photocurrent from the well.

While for this qualitative discussion, an oversimplistic band structure model
was used, a more accurate, atomistic approach is required to quantitatively model
carrier escape, taking into account the non-parabolic and anisotropic
transverse dispersion of the confinement levels at all energies in the well.
For a proper description of carrier heating, confinement effects should also be
considered for the phonons. While this gives a more accurate picture of the
carrier escape, it is still necessary to consider on the same level also the
recombination mechanisms that compete with it, i.e. extend the discussion to
nonradiative recombination, in order to obtain a realistic internal quantum
efficiency determining the effectiveness of the system in photovoltaic
applications.

\section*{Acknowledgements}
The author acknowledges financial support from the Institute of Theoretical
Physics at ETH Zurich during the initial stage of this work.

\section*{References}
\bibliographystyle{elsarticle-num}

\end{document}